\documentclass{optica-article}

\journal{opticajournal} 

\articletype{Research Article}

\usepackage{lineno}
\hypersetup{
    colorlinks=true,
    linkcolor=blue,
    filecolor=blue,      
    urlcolor=blue,
    citecolor=blue
}

\begin{document}

\title{Quantum nonlocal double slit interference with partially coherent qubits}

\author{Sakshi Rao and Bhaskar Kanseri\authormark{*}}

\address{Experimental Quantum Interferometry and Polarization (EQUIP), Department of Physics, Indian Institute of Technology Delhi, Hauz Khas, New Delhi 110016, India}

\email{\authormark{*}bkanseri@physics.iitd.ac.in} 


\begin{abstract*} 
Partially coherent quantum-entangled beams combine quantum entanglement with partial coherence, allowing them to maintain quantum characteristics while being more resistant to distortions caused by random media during propagation. In this study, we investigate the effect of coherence variation of such beams on non-local double-slit quantum interference. The spatial coherence variation is achieved by controlling the spot size and transverse coherence length of the Gaussian Schell model pump in the spontaneous parametric down conversion process. For a fixed beam size, the momentum correlation width of partially coherent biphotons increases with the decreases in the transverse coherence length. This results in a biphoton beam exhibiting multiple spatial modes, making it more suitable for studying the non-local features of quantum states in imaging, interference, and diffraction experiments. Our findings infer both high-quality and near-unity visibility of nonlocal interference using the partially coherent twin beams, even with the substantial decrease in the coherence of the pump. We believe these results can enhance robustness against the deleterious effects of the medium during propagation and can have potential applications in optical image cryptography, biomedical imaging, quantum lithography, and quantum holography.

\end{abstract*}

\section{Introduction}
Our understanding of quantum processes has been aided by research in quantum optics by two-photon coincidence rate measurements through interference from nonlocal double-slit, ghost imaging experiments by manifesting the nonlocal features of the entangled photon source\cite{shih2024ghost,aspden2013epr,ribeiro1996direct,fonseca1999quantum,d2005quantum}.
Pioneer work demonstrates how photon pairs produced by the spontaneous parametric down-conversion (SPDC) process\cite{rubin1996transverse,couteau2018spontaneous} create an image by exploiting their spatial entanglement in both near- and far-fields. In nonlocal double-slit interference, the interference pattern is created after scattering the biphotons from two spatially separated different apertures, emphasizing the role played by spatially correlated photon pairs using the fourth-order correlation measurement. The superposition of these two spatially separated apertures forms an interference pattern without directly using a double-slit in any path.  The local interpretation of this type of experiment is not possible, since the formed pattern is not derived from the direct intensity measurement of the signal and the idler beam. This phenomenon offers potential applications in ghost imaging, quantum holography, quantum lithography, super-resolution imaging techniques, and optical cryptography. Later, this non-local double-slit effect has been shown in various theoretical and experimental studies using both classical and quantum light\cite{gao2008interference,gan2009interference, walborn2010spatial,jost1998spatial,boucher2021engineering}. 
\\
\\
Building on this foundation, \cite{gan2009interference} explored interference using a spatially incoherent light source illuminating two separated apertures. Their work attributed the observed interference effects using intensity measurement.  In addition to spatially incoherent sources, partially coherent beams with a tunable degree of coherence(DoC) can provide enhanced flexibility for light manipulation and thus introduce unique features and novel applications in the field of quantum free space communication\cite{cai2017generation,kanseri2020development,shirai2003mode,liu2014experimental,ricklin2002atmospheric,wang2015propagation}. These beams possess self-healing properties through propagation, which improves spatial profiles and polarization in the turbulent media and opens up new possibilities for secure communication and  sensing\cite{zhou2020application,kim2005optical}.
These pseudothermal or partially coherent light sources have been seen to be very useful in ghost interference\cite{vidal2009effects,luo2015ghost} and ghost imaging \cite{ferri2005high,cai2004ghost}.
The quality and visibility of ghost images can be controlled by adjusting the transverse size and coherence width of the source. However, achieving both good visibility and high resolution simultaneously is challenging in the classical case because of the intrinsic background noise term in the fourth-order correlation function.
\\
\\
In contrast, if we consider ghost interference and nonlocal interference with EPR-correlated biphotons, background-free images with high visibility and better resolution have been achieved simultaneously in both position and momentum anti-correlated configuration \cite{aspden2013epr}. These spatially entangled photon pairs are the basis of many exciting fields such as quantum cryptography\cite{ekert1992}, quantum free space communication\cite{Aspelmeyer2003}, and quantum metrology and sensing applications\cite{polino2020,pirandola2018advances,wang2021classicalsensing}. Quantum entangled beams combined with partial coherence features, resulting in partially coherent qubits exhibit both quantum and partial coherence properties, offering enormous potential for a variety of applications such as in free space communication, and in precision measurement\cite{sharma2021controlling, phehlukwayo2020influence, rao2022investigation}.  By tailoring the pump parameters, the spatial correlation of the biphotons can be controlled \cite{Saleh2000, jha2010spatial, defienne2019spatially}. More recently, it has been shown theoretically and experimentally that the partially coherent features of the pump beam transfer to the quantum correlated beams through the SPDC process, as achieved using a Gaussian Schell model (GSM) pump source \cite{sharma2023experimental,sharma2021controlling,rao2024recovery}. The effect of partial coherence introduced in the path of biphotons to enhance the quality of ghost imaging has been explored in some works \cite{hong2018two,lib2022thermal}.  However, non-local interference and ghost imaging with partial coherent quantum beams generated with GSM pumped SPDC has not yet been investigated so far. Owing to their intrinsic multi-mode nature, the partially spatially coherent qubits can overcome the difficulties caused by atmospheric turbulence\cite{qiu2012influence, qiu_she2012,lucas2020} and potentially improve quantum holography, secure communication, and biological imaging.

In this study, we introduce a method to examine how the transverse size and coherence length of the pump influence nonlocal double-slit interference involving partially coherent qubits. Our approach specifically focuses on using a GSM pump source with tunable spatial coherence. This approach provides a theoretical framework for manipulating the coherence properties of the beam, allowing precise control over its spatial coherence characteristics. By incorporating this tunable GSM pump into the SPDC process, we can effectively regulate the multiple spatial mode nature of the generated biphotons and explore their impact on quantum imaging.
  For the detection part, we replace the traditional two single-photon detectors with a single electron-multiplying charge-coupled device (EMCCD). In the camera plane, we can precisely position the apertures at the correlation points within the biphoton spatial profile. This setup enables efficient capture of multiple spatial modes in a single shot measurement. By tuning the spatial coherence of the GSM pump, we systematically investigate how variations in coherence influence the interference pattern observed in coincidence measurements.

The structure of the paper is organized as follows: 
In section II, we discuss the theoretical framework to study the effect of spatial correlation of partially coherent qubits generated with a GSM pump on the nonlocal double-slit interference pattern. We derive an expression of the coincidence count rate of the partially coherent qubits at the detection plane after scattering through the two separated apertures. In section III, we discuss the generation and characterization of a partially spatially coherent pump. We outline the experimental scheme for the nonlocal double-slit interference using the partially spatially coherent qubits.
Section IV includes the results and discussion, focusing on the impact of partial coherence of the spatially correlated qubits on the quality and visibility of the interference pattern. Finally, Section V concludes the study. 

\section{Methodology}
The scheme for nonlocal quantum double-slit interference with partially coherent qubits is shown in Fig.1. A GSM beam is used to pump a type-II nonlinear crystal. The aperture $ A_1$ (single slit) in the signal beam and $A_2$ (wire) in the idler beam are positioned at the same distance $z_0$ from the crystal. Coincidence detection between the two detectors, located at an equal distance $z_1$ from each aperture, can create a nonlocal double-slit interference pattern.
The quantum state generated by the SPDC process using a GSM beam is given byby\cite{phehlukwayo2020influence}.
\begin{figure}
\centering
 \includegraphics[width=0.7\columnwidth]{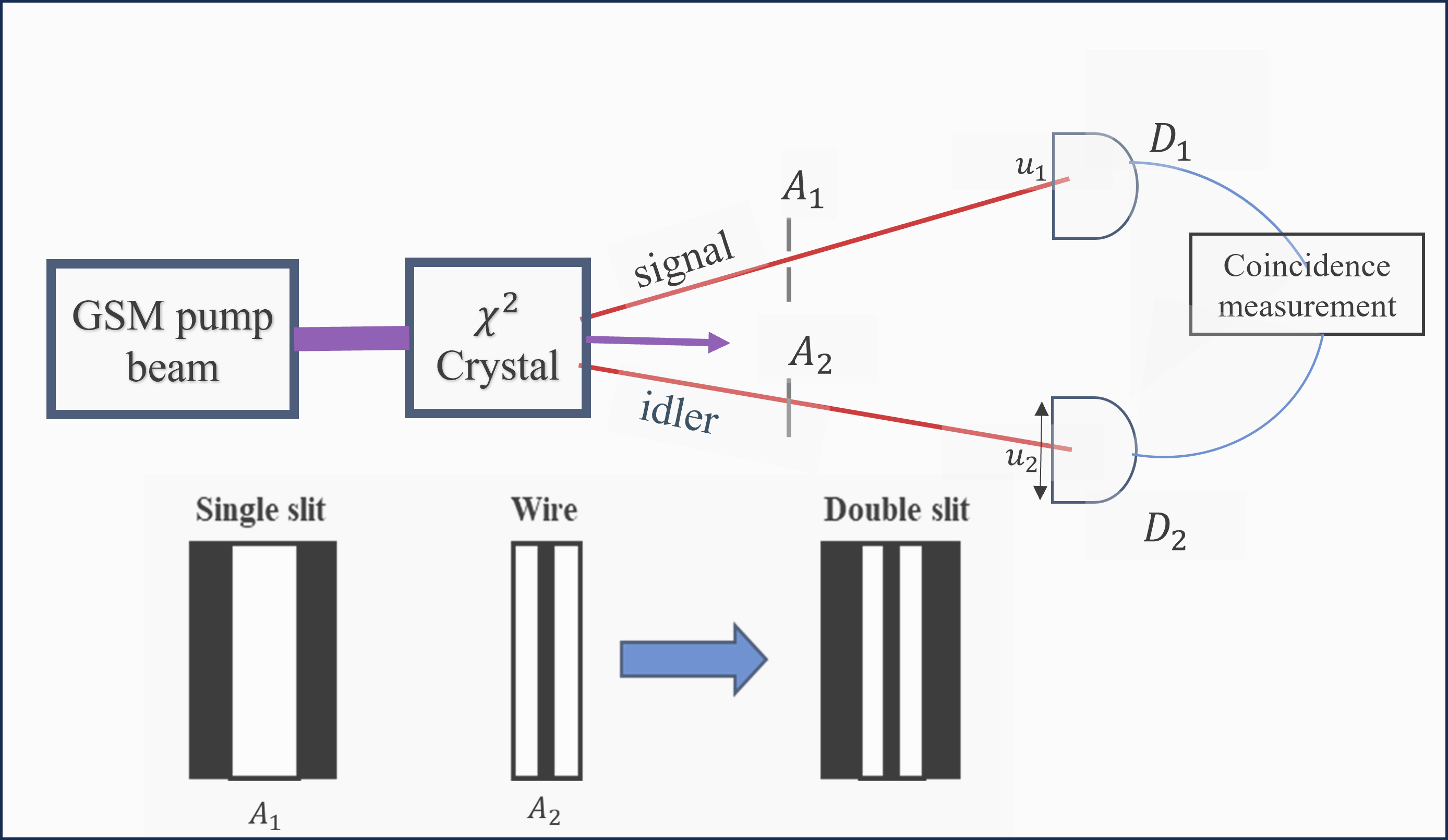}
    \caption{Scheme for nonlocal interference with partially coherent qubits.$A_{1}$, $A_{2}$ are single slit and wire respectively and $D_{1}$ and $D_{2}$ are detectors.}
    \label{fig:setup}
\end{figure}
\begin{equation}
   |\Psi \rangle=\iint d\mathbf{k}_{i} d\mathbf{k}_{s} \Phi (\mathbf{k}_{i}, \mathbf{k}_{s}) a^{+}_{1}(\mathbf{k}_{i})a^{+}_{2}(\mathbf{k}_{s})|0,0\rangle.
\end{equation}
Here, $\Phi (\mathbf{k}_{i}, \mathbf{k}_{s})$ is the transverse profile of the twin beam with $|0,0\rangle$ as the vacuum state. The creation and annihilation operator are given by $a^{+}$ and $a$, respectively.
The two-photon wave packets for idler and signal photons of orthogonal polarization have been generated in non-collinear type-II SPDC configuration with the pump modeled as a Gaussian Schell model beam. The degree of spatial coherence of the pump is given by $A = \frac{\delta}{2 w_0}$, where $\delta$ denotes the effective spectral width, defined as
$\delta = \sqrt{\frac{1}{\left( \frac{1}{l_c^2} + \frac{1}{4w_0^2} \right)}}$
with $w_0$ representing the beam size and $l_{c}$ is the transverse coherence length of the pump\cite{sharma2021controlling}.
\begin{figure*}
    \centering
    \includegraphics[width=0.95\textwidth]{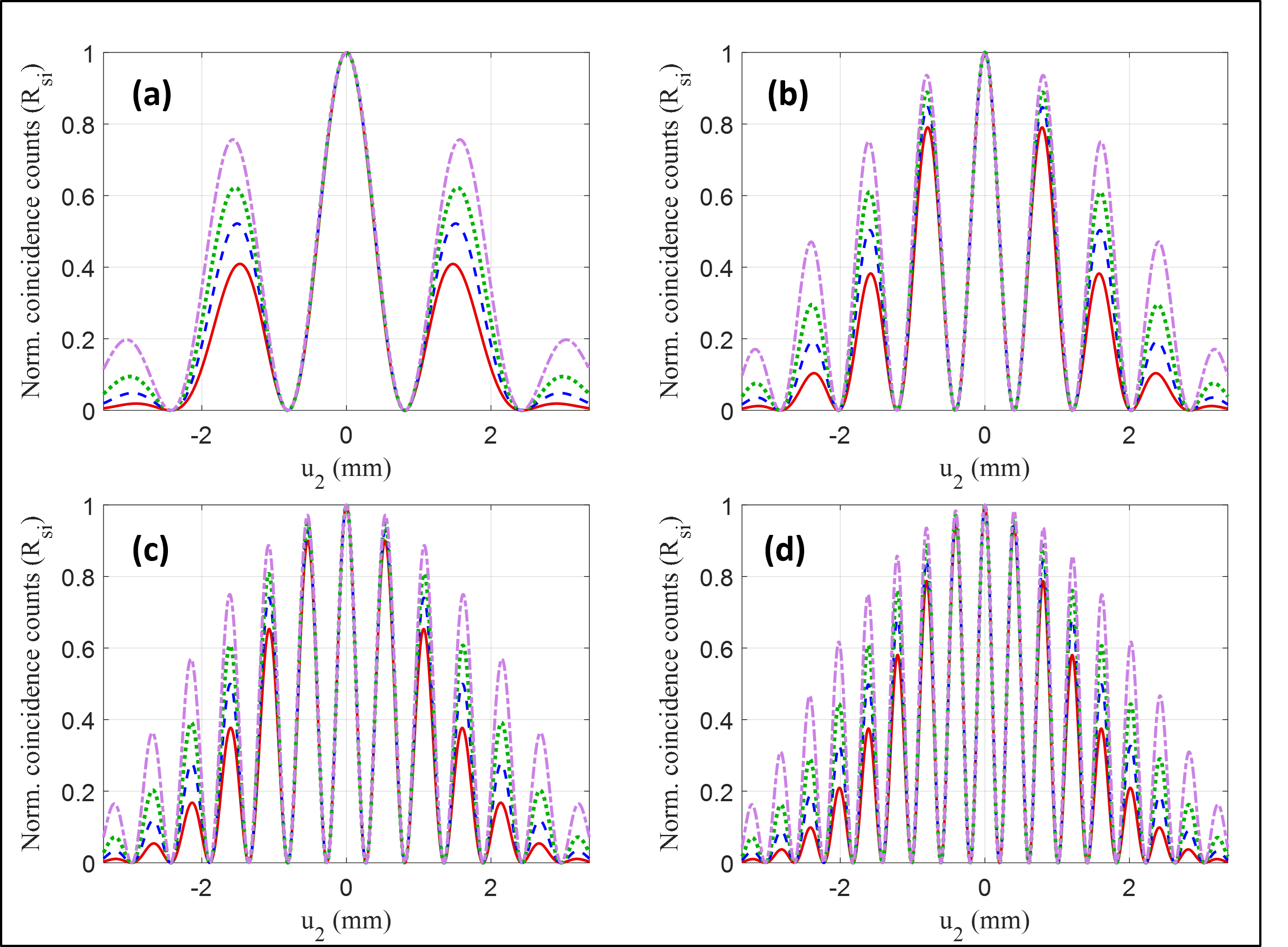}
   \caption{Interference pattern of the nonlocal double slit with partially spatially coherent qubits for different degree of spatial coherence, red, $A=0.99$; blue, $A=0.76$; green, $A=0.5$; purple, $A=0.2$. The wire of diameter $a=80\mu m$ with different silt width $b$. (a) for $b=0.2mm$; (b) for $b=0.4mm$; (c) for $b=0.6mm$; and (d) for $b=0.8mm$.}
    \label{theory}
\end{figure*}
 Spatial coherence can be quantified using the parameter $A$, where fully spatially incoherent beams have a value of 0, and fully spatially coherent beams have a value of 1. Beams with partial spatial coherence exhibit intermediate $A$ values within 0 and 1.
\\
The probability of coincidence count rate used to express the distribution of these transmitted idler and signal photons aimed at the camera is given by\cite{phehlukwayo2020influence}.
\begin{equation}
R(u_{1},u_{2})=\langle {\Psi}|\hat{E}^{-}_{s}(\vec{\rho},z) \hat{E}^{-}_{i}(\vec{\rho},z)\hat{E}^{+}_{s}(\vec{\rho},z)\hat{E}^{+}_{i}(\vec{\rho},z)|\Psi\rangle.
\end{equation}
Here, $u_{1}$ and $u_{2}$ are the positions of the detectors $D_{1}$ and $D_{2}$, respectively.
The above expression can also be written as
\begin{equation}
\begin{split}
R_{si}(u_{1},u_{2})=&\iiiint
     W^{2}{(\rho_{s},\rho_{i},\rho'_{s},\rho'_{i},\theta_{s},\theta_{i},z})\\
     & \times
 h_{1}(u_{1},\rho_{s}) h_{2}(u_{2},\rho_{i})  h_{1}^{*}(u_{1},\rho'_{s}) h_{2}^{*}(u_{2},\rho'_{i}) d{\rho_{s}} d{\rho_{i}} d{\rho'_{s}} d{\rho'_{i}},
    \end{split}
\end{equation}
with the cross-spectral density function of partially coherent spatially entangled qubits generated through the SPDC process using a GSM beam given by the expression as\cite{jha2010spatial}
\begin{equation}
\begin{split}
&W^{(2)}({\rho}_{s},{\rho}_{i},{\rho'}_{s},{\rho'}_{i},\theta_{s},\theta_{i},z)\\&=A_{01}\exp\Big[\frac{\iota k_{p}}{4z}({\rho}_{s}^2+{\rho}_{i}^2-{\rho'}_{s}^2-{\rho'}_{i}^2\Big] \exp\Big[\frac{-(2w_{0}^{2}+l_{c}^{2})\delta^{2}k_{p}^{2}}{16z^{2}l_{c}^{2}}\big[({\rho}_{s}+{\rho}_{i})^{2}+({\rho'}_{s}+{\rho'}_{i})^{2}\big]\Big]\\
&\times
\exp\Big[\frac{-w_{0}^{2}\delta^{2}k_{p}^{2}}{4z^{2}l_{c}^{2}}\big[({\rho}_{s}+{\rho}_{i})({\rho'}_{s}+{\rho'}_{i})\big]\Big]\sin(2\theta_{i})\sin(2\theta_{s}),
\end{split}
\end{equation}
where, $\rho_{s}$, $\rho_{i}$, $\rho_{s}$, $\rho'_{s}$, $\rho'_{i}$ are transverse positions of signal and idler photons. 
$l_{c}$, $w_{0}$, and $k_{p}$ denote the transverse coherence length, beam size, and wave vector of the GSM pump, respectively.

In the provided Eq.3, the function $h_{j}(u_{j},\rho)$  denote the impulse response function for the two optical paths, taking on the following forms
\begin{equation}
h_{j}( {u_{j}},{\rho})=\frac{1}{\lambda^{2}z_{0}z_{1}}\int\exp\Big[\frac{-\iota k}{2z_{0}}({\rho}-{v}_{j})^{2}\Big] exp\Big[\frac{-\iota k}{2z_{1}}({v}_{j}-{u}_{j})^{2}\Big] A_{j}({v_{j}})d{v_{j}}.
\end{equation}
Here, $j=s$ and $i$, $A(v_{1})=rect(\frac{v}{a})$ and $A(v_{2})=1-rect(\frac{v}{b})$ represent the transmission function of the two apertures.
By incorporating all these expressions into Eq. 3, we obtain the expression given by

\begin{equation}
\begin{split}
R_{si}(u_{1},u_{2})=&D_{0}\exp(\frac{k^2 (u_1^2 + u_2^2)}{2 z_1^2 e}) 
\exp( \frac{f^2}{4 l}) 
\exp( \frac{b_1^2}{4 A_0}) 
\exp( \frac{n^2}{4 m} )
\exp( \frac{t^2}{4 s} )\\
&\times \big[\operatorname{sinc}( k (u_1 - u_2) a / z_1)
\cos( k (u_1 - u_2) b / z_1)\big]^2,
\end{split}
\end{equation}
where,
\begin{equation}
\begin{aligned}
e &= \frac{i k}{2 z_0} + \frac{i k}{2 z_1}; \quad B_0= \frac{k_p^2 l_c^2 (w_0^2 + 2 l_c^2)}{4 z^2 (w_0^2 + 2 l_c^2)};
A_0= -\frac{i k_p}{z} + \frac{i k}{z} - B_0 + \frac{k^2}{4 e z_0^2};  \quad b_2 = \frac{k^2 u_2}{2 e z_0 z_1};\\
\quad D_{0}&= \frac{1}{\lambda^4 z_0^2 z_1^2} \frac{\pi^2}{\sqrt{A_{0}sme}}; \quad b_1= \frac{k^2 u_1}{2 e z_0 z_1}; 
\quad A_2= -\frac{i k_p}{z} + \frac{i k}{z} - B_0 - \frac{k^2}{4 e z_0^2}; \quad f= b_2 + \frac{B_0}{A_0};\\
g &= C_0 \left(1 + \frac{B_0 C_0}{A_0}\right);
\quad m = A_2 - \frac{C_0^2}{4 A_0} - \frac{B_0^2}{l} - \frac{g^2}{4 l} + \frac{g B_0}{l};  \quad l = A_0 - \frac{B_0^2}{A_0};\\
n &= b_1 + \frac{C_0 b_1}{2 A_0} - \frac{B_0}{l} + \frac{g f}{2 l};\quad p= -\frac{g^2}{2 l} + \frac{g B_0}{l} - \frac{C_0}{2 A_0}; \quad C_0= \frac{4 w_0^2 l_c^2 k_p^2}{2 z^2 (w_0^2 + 4 l_c^2)}; \\
s &= A_2 - \frac{C_0^2}{4 A_0} - \frac{p^2}{4 m} - \frac{g^2}{4 l};
\quad t= \frac{C_0 b_1}{2 A_0} + b_2 + \frac{g f}{2 l} - \frac{n p}{2 m}.
\end{aligned}
\end{equation}
Here, $k=k_{s}=k_{i}=2\pi/\lambda_{si}$ and $k_{p}=2\pi/\lambda_{p}$, with wavelength of signal and idler beam denoted by $\lambda_{si}=810nm$ and pump beam is denoted by $\lambda_{p}=405nm$. 
To reconstruct the double-slit pattern, fourth-order correlation measurements from two detectors are conducted by scanning the detector in the transverse direction. The transfer of the angular spectrum of the pump used to generate spatially entangled states allows us to control the fourth-order correlation function\cite{ribeiro1996direct,edgar2012imaging}. By manipulating the pump beam, we can influence the coincidence count rate, leading to variations in the visibility and quality of the interference pattern. 
\\
The visibility of two-photon interference is given by 
\begin{equation}
    V=\frac{R_{si}(u_{1},u_{2})_{\text{max}}-R_{si}(u_{1},u_{2})_{\text{min}}}{R_{si}(u_{1},u_{2})_{\text{max}}+R_{si}(u_{1},u_{2})_{\text{min}}}.
\end{equation}
 
Fig.2 shows the normalized coincidence count rates of entangled photons at positions $u_{1}=0$ and $u_{2}$ (scanning), which shows the joint
 detection of the entangled photons on the position of
 $u_{1}=0$ and $u_{2}$ (scanning) described by the fourth-order
 correlation function. The two-photon coincidence probability for any given arbitrary quantum state of light is proportional to the intensity correlation function, which is provided by\cite{saleh2005wolf}.
 In these plots, we varied the single slit width while keeping the wire width fixed at $b=80\mu m$. The increase in single slit width leads to more visible fringes in the same total width of the pattern, and our results indicate that as the degree of spatial coherence of the pump decreases, the resolution of the fringes improves and thus making them more helpful in gathering the more spatial details about the apertures. 
\section{Experimental Setup}
 The experimental setup for generating and characterizing the partially spatially coherent pump is depicted in Fig. 3(a). The laser beam with a wavelength of $405nm$  is directed through lens $L_1$  having a focal length of $40 mm$, which focuses the beam onto the rotating ground glass diffuser (RGGD). The RGGD scatters the light to create an incoherent beam. Adjusting the position of $L_1$, changes the spot size on the diffuser, allowing control over the degree of spatial coherence. Owing to the van Cittert-Zernike theorem, a partially coherent beam is obtained using the ground glass diffuser at the back focal plane of a lens $L_{2}$ with a focal length of $200 mm$. The output collimated light is a partially coherent beam whose spatial coherence varies with the spot size of the beam at the diffuser. The partially coherent beam is then passed through a Gaussian amplitude filter (GAF), and a Gaussian Schell model beam is formed. We calculated the visibility of fringes formed using four double slits with a fixed slit width of $ 0.15 mm$, and different slit separation $0.25 mm, 0.5 mm, 0.75 mm$ and $1 mm$ respectively with the expression $ \text{Visibility}=|\gamma ({\bf r}_{1},{\bf r}_{2},\tau )|= \Big| 2 \frac{J_{1}(\nu )}{\nu }\Big|$\cite{sharma2021controlling}, where $\gamma ({\bf r}_{1},{\bf r}_{2},\tau )$ represents the degree of spatial coherence. The parameter $\nu$ is defined as $\nu = \frac{k_{p} d_{12} a_{s}}{f}$ with $k_{p}=\frac{2 \pi}{\lambda_{p}}$ as wavevector of the beam, $d_{12}$ is slit separation, $a_{s}$ is beam radius at surface of diffuser and $f$ is the focal length of lens $L_{2}$.
\begin{figure*}
    \centering
    \includegraphics[width=\textwidth]{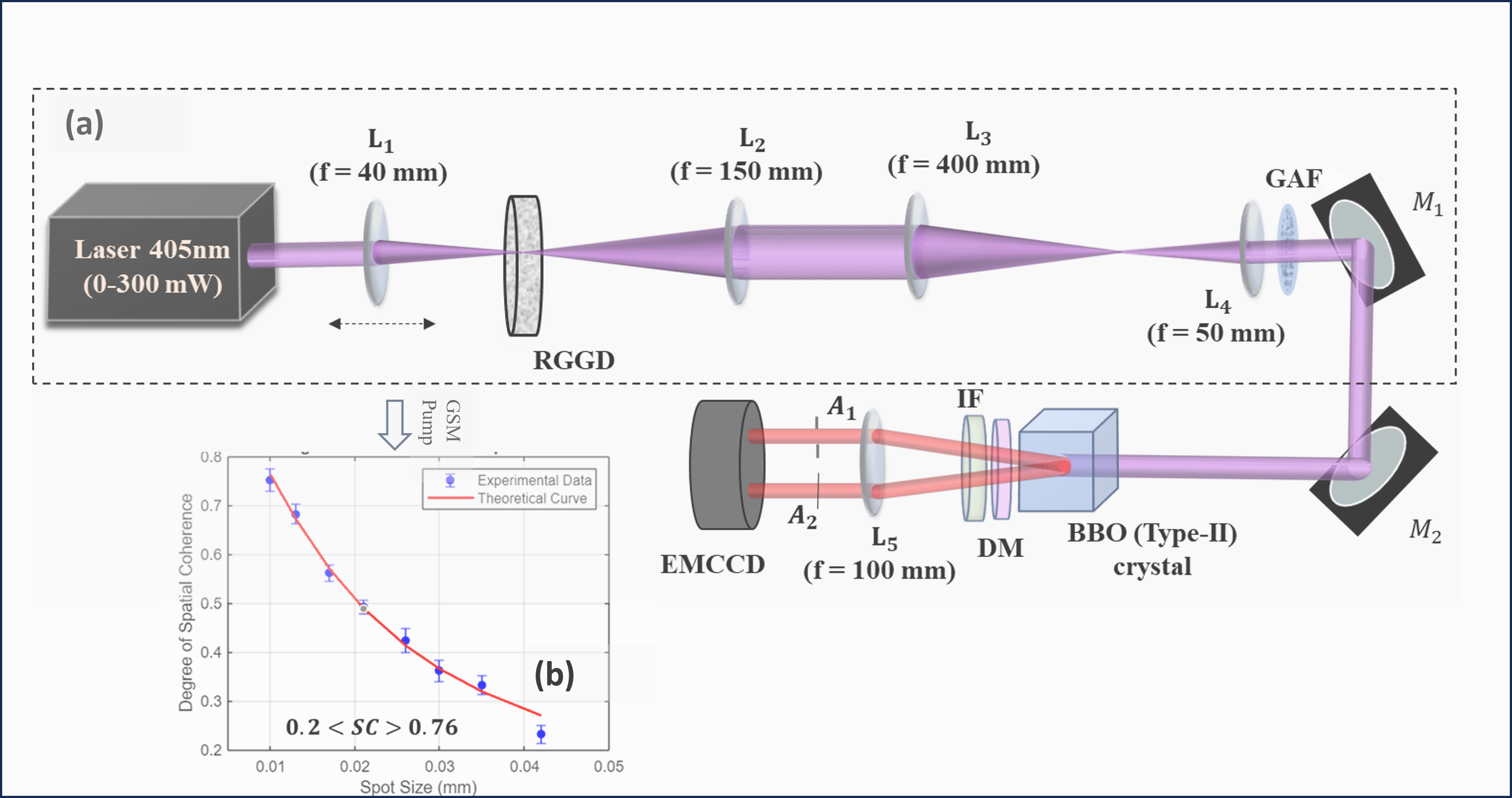}
    \caption{(a)Experimental setup for nonlocal interference from a wire and single slit in the path of spatially entangled photons generated with Gaussian Schell model (GSM) beam (b) to measure the degree of spatial coherence of the GSM pump.Notations: L: lens; RGGD: rotating ground glass diffuser; GAF: Gaussian
 amplitude filter; BBO: beta barium borate; DM:dichroic mirror; IF:interference filter; SC:spatial coherence.}
    \label{experiment_setup}
\end{figure*}
 By fitting the visibility, we estimated the spot size at the RGGD with the translation of the lens $L_{1}$ and calculated the transverse correlation length as \( l_{c} = \frac{3.832 f}{k_{p} a_{s}} \) \cite{mandel1995optical}.  We created a partially coherent source with a tunable degree of coherence ranging from $A=0.76$ to $A=0.2$ by varying the spot size at the diffuser.

The size of the beam is then reduced by the factor of $8$ using lenses $L_{3}$ and $L_{4}$ with focal lengths $400 mm$ and $50 mm$, respectively. The pump beam is then incident on the second-order non-linear crystal to generate the down-converted photons. Pump beam size should essentially remain constant while examining the impact of variations in the coherence length of the pump on biphotons. The beam size at the crystal position is checked by placing a beam profiler and found to be $w_{0}=2.3mm$, which is almost constant with the translation of the lens $L_{1}$.
 We have examined the variation in spatial coherence indicated by the pump beam parameter ($A$) due to the lens translation ($L_1$), which alters the spot size at the diffuser. The plot shown in Fig.3(b) illustrates the change in the degree of spatial coherence with the variation of spot size at the RGGD\cite{kanseri2020development}.
\\
\begin{figure*}
    \centering
    \includegraphics[width=\textwidth]{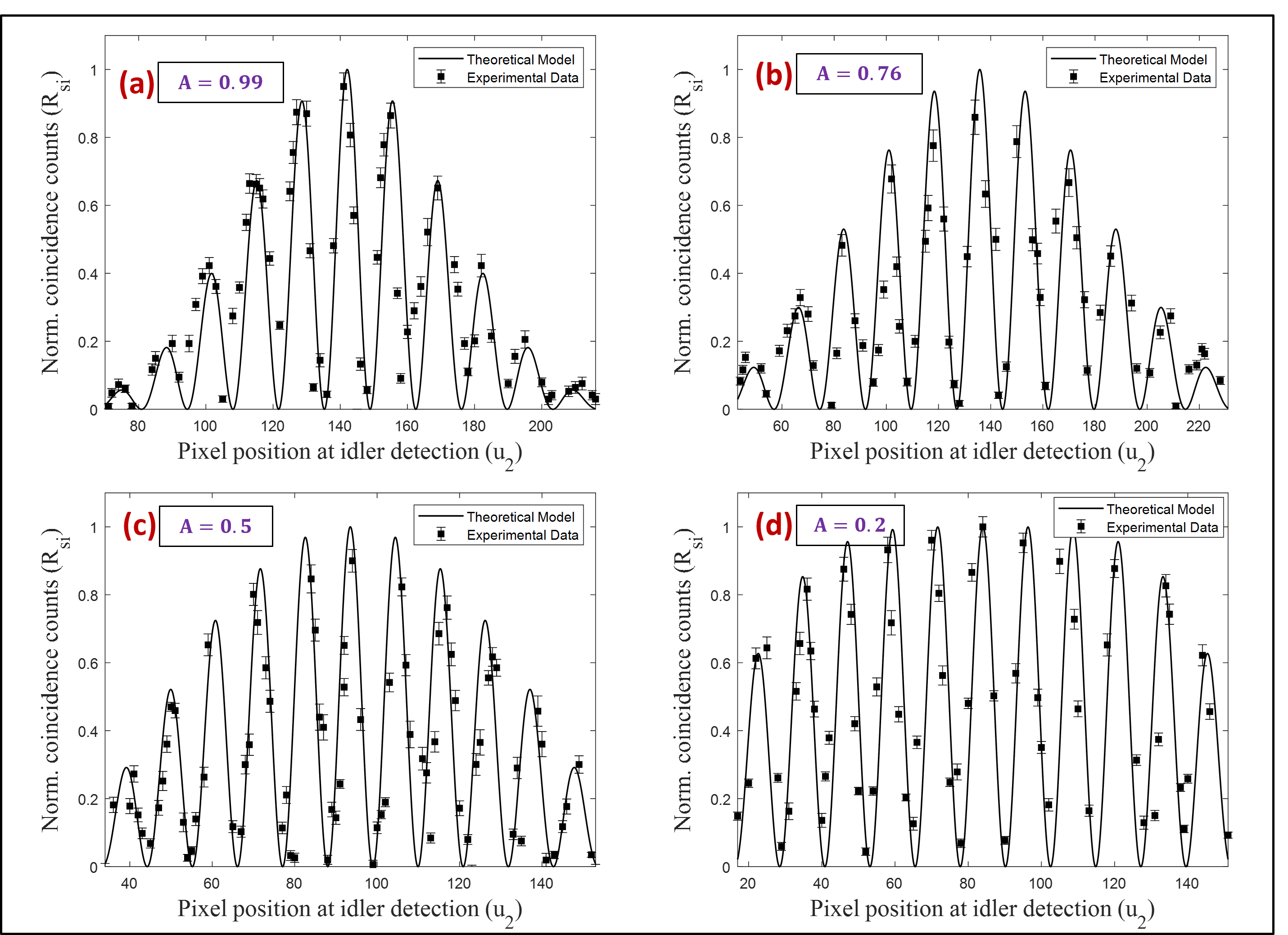}
   \caption{Plot of normalized coincidence counts as a function of the pixel position at aperture $A_{1}$
  on the detector, with the pixel position at $A_{2}$ fixed for four different cases of degree of spatial coherence. (a) for  $A=0.99$; (b) for $A=0.76$; (c) for $A=0.5$; (d) for $A=0.2$ with fixed $w_0=2.3mm$.}
    \label{experimental_plot}
\end{figure*}
We used a type II noncollinear degenerate SPDC using a beta barium borate (BBO) crystal as the down-converter, which generates spatially correlated photon pairs. 
The apertures $A_{1}$ and $A_{2}$ are strategically positioned in the marked conjugate positions of the signal and idler photons, where $A_{1}$ is a single slit of width $d=0.55mm$ and $A_{2}$ is a wire of width $a=80\mu m$. For the detection, we used an EMCCD camera, and the coincidence probability was calculated by image processing. The EM gain was set at 1000, with a read-out rate of $17$ MHz horizontal pixel shift and a vertical shift $0.3 \mu s$. The vertical clock amplitude voltage was set to $+4$ V, and the camera temperature was $-80$ C. The coincidence counts between $i$ and $j$ pixel position in captured images are calculated using $C=\langle n_{i}n_{j} \rangle -\langle n_{i}\rangle \langle n_{j}\rangle $ \cite{Avella2016} where $n_{i}$ and $n_{j}$ are average counts at position $i$ and $j$, respectively, for the position $u_{1}$ and $u_{2}$ of the apertures. For each setting of the degree of spatial coherence of the pump $A=1,0.76,0.5,0.2$, we captured 40,000 frames with an exposure time $2.5ms$ by operating the EMCCD in photon counting mode to calculate coincidence rates, which is directly proportional to the fourth-order correlation function, providing insights into the quantum properties of the generated photon pairs.
\section{Result and Discussion}
The interference pattern arises from the two spatially separated apertures with partially coherent qubits using their momentum correlation in the absence of the physical double slit. The coincidence rate is calculated as a function of the marked positions of the apertures $A_{1}$ and $A_{2}$. The experimental setup allows for investigating the partially coherent features inherent in the spatially correlated photon pairs produced through SPDC when pumped with a GSM beam. This contributes to a deeper understanding of the nonlocal correlations between the photons and highlights the potential of these partially coherent qubits for applications in quantum imaging and sensing and fundamental studies of quantum nonlocality.
\begin{figure}
    \centering
    \includegraphics[width=0.7\columnwidth]{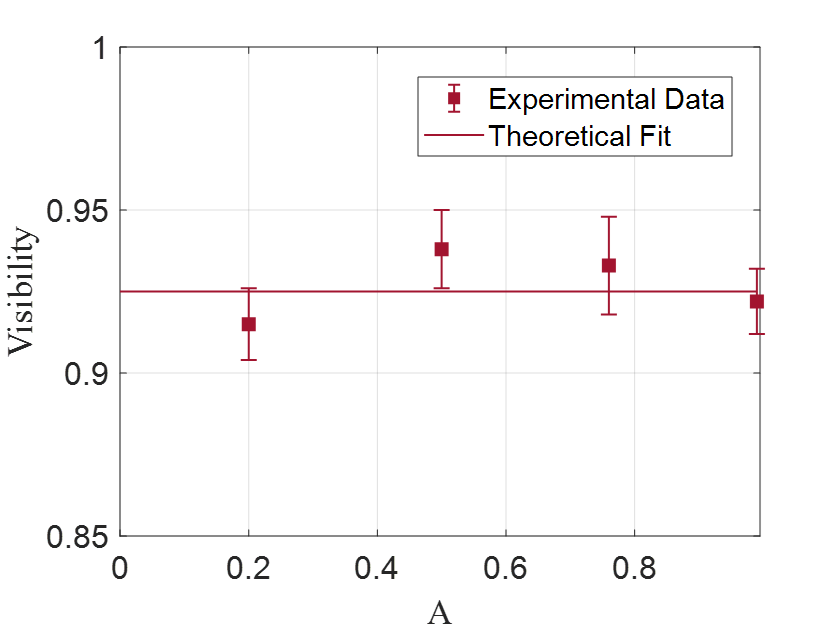}
  \caption{Normalized visibility vs degree of spatial coherence of the pump.}
    \label{visibility}
\end{figure}
Fig.4 shows the plot of the normalized coincidence count as a function of the pixel position translation on the side where aperture $A_{1}$ is located, while the pixel position of the aperture $A_{2}$ remains fixed. The quality of interference fringes remains high with the decrease in the spatial coherence of the pump, demonstrating that even with partially coherent qubits, we can achieve better-quality images. This suggests that more information about the spatial structure of the aperture or object can be extracted, which can be particularly useful in ghost imaging and quantum holography. Fig.5 shows the plot of experimental visibility with the variation in the degree of spatial coherence of the GSM pump beam. The visibility is almost constant for all cases of spatial coherence of the GSM pump and shows a high value even with a lower degree of spatial coherence. 

Numerous quantum imaging approaches, such as ghost imaging\cite{aspden2013epr}, sub-shot-noise imaging\cite{ruo2020improving}, and sub-Rayleigh imaging\cite{xu2015experimental}, rely on the properties of the down-converted photons and are based on spatially entangled photon-pair sources. However, by altering the coherence characteristics of the pump, the spatial structure of the two-photon wave function and, consequently, its joint probability distribution can be changed. Adjusting beam size and coherence length of the GSM pump improves image quality and can be more advantageous than the SPDC photons generated by a fully coherent laser. Its coherence length $l_{c}$ and pump beam size $w_{0}$ determine the momentum correlation width $w_{k}=\sqrt{1/l_c^2+1/w_{0}^2}$. For a fixed beam size, this expression indicates that as the coherence length of the GSM pump decreases, the momentum correlation width increases. This broader momentum correlation increases multiple spatial modes, making partially coherent qubits more versatile. As a result, they can significantly improve the resolution and clarity of applications such as ghost imaging, quantum holography, and optical encryption, particularly in complex or turbulent media. By leveraging this property, partially coherent qubits could enhance the retrieval of spatial information, reduce noise, and improve robustness in quantum imaging and communication systems. These partially coherent qubit beams can have an advantage over classical beams, particularly in biomedical imaging of photosensitive objects such as live cells and biomedical tissues with multiple scattering levels, in transferring encrypted information through free space in the turbulent medium, and in quantum holography and ghost imaging to get better spatial information of the object.
\section{Conclusion}
In summary, we investigated the interference of partially spatially correlated entangled qubits formed after passing through the nonlocal double-slit experiment. By varying the spatial coherence of the pump in the type-II SPDC process, we analyze the impact of spatial correlations on the quality and visibility of the interference pattern theoretically and experimentally.
The experimental results align well with theoretical predictions and show that the quality of the interference pattern remains high with the decrease in the degree of coherence of the pump.
Additionally, high visibility is maintained for partially spatially coherent qubits, enabling us to achieve both high visibility and superior quality simultaneously. 
The momentum correlation width increases with a decrease in the degree of spatial coherence of the pump, which makes these photons more robust while propagating through the distorted medium by mitigating the speckle noise and improving the clarity of reconstructed phase or amplitude images.
These results could be highly beneficial for quantum metrology and sensing experiments, providing new pathways to improve the precision and reliability of quantum-based technologies.
\section{Funding}
 Author SR acknowledges financial support from DST INSPIRE for a Senior Research Fellowship (SRF). 
\section{Disclosures}
The authors declare no conflicts of interest.
\section{Data availability}
 The data supporting the findings of this paper are not publicly accessible at this time but can be requested from the authors under reasonable conditions.

\end{document}